\documentclass[pra,twocolumn]{revtex4-1}
\usepackage{graphicx}
\usepackage{amssymb,amsmath}
\usepackage{ulem,color}
\newcommand{\ket}[1]{\left|#1 \right\rangle}
\newcommand{\bra}[1]{\left\langle #1 \right |}

\newcommand{\bbsout}[1]{}

\newcommand{\bsout}[1]{}
\usepackage{siunitx}
\usepackage[english]{babel}
\usepackage{blindtext}
\usepackage{hyperref}
\usepackage[T1]{fontenc}
\usepackage{mwe}

\begin{document}
\title{Quantum enhancement of signal-to-noise ratio with a heralded linear amplifier}

\author{Jie Zhao}
\affiliation{Centre for Quantum Computation and Communication Technology, Research School of Physics and Engineering, Australian National University, Canberra ACT 2601, Australia.}

\author{Josephine Dias}
\affiliation{Centre for Quantum Computation and Communication
Technology, School of Mathematics and Physics, University of
Queensland, St. Lucia, Queensland 4072, Australia.}

\author{Jing Yan Haw}
\affiliation{Centre for Quantum Computation and Communication Technology, Research School of Physics and Engineering, Australian National University, Canberra ACT 2601, Australia.}

\author{Mark Bradshaw}
\affiliation{Centre for Quantum Computation and Communication Technology, Research School of Physics and Engineering, Australian National University, Canberra ACT 2601, Australia.}

\author{R$\acute{e}$mi Blandino}
\affiliation{Centre for Quantum Computation and Communication
 Technology, School of Mathematics and Physics, University of
Queensland, St. Lucia, Queensland 4072, Australia.}

\author{Thomas Symul}
\affiliation{Centre for Quantum Computation and Communication Technology, Research School of Physics and Engineering, Australian National University, Canberra ACT 2601, Australia.}

\author{Timothy C. Ralph}
\affiliation{Centre for Quantum Computation and Communication
 Technology, School of Mathematics and Physics, University of
Queensland, St. Lucia, Queensland 4072, Australia.}

\author{Syed M Assad}%
  \email{cqtsma@gmail.com}
\affiliation{Centre for Quantum Computation and Communication Technology, Research School of Physics and Engineering, Australian National University, Canberra ACT 2601, Australia.}

 \author{Ping Koy Lam}
\affiliation{Centre for Quantum Computation and Communication Technology, Research School of Physics and Engineering, Australian National University, Canberra ACT 2601, Australia.}

\begin{abstract}
Due to the pervasive nature of decoherence, protection of quantum information during transmission is of critical importance for any quantum network. A linear amplifier that can enhance quantum signals stronger than their associated noise while preserving quantum coherence is therefore of great use. This seemingly unphysical amplifier property is achievable for a class of probabilistic amplifiers that does not work deterministically. Here we present a linear amplification scheme that realises this property for coherent states by combining a heralded measurement-based noiseless linear amplifier and a deterministic linear amplifier. The concatenation of two amplifiers introduces the flexibility that allows one to tune between the regimes of high-gain or high noise-reduction, and control the trade-off of these performances against a finite heralding probability. We demonstrate an amplification signal transfer coefficient of $\mathcal{T}_s > 1$ with no statistical distortion of the output state. By partially relaxing the demand of output Gaussianity, we can obtain further improvement to achieve a $\mathcal{T}_s = 2.55 \pm 0.08$. Our amplification scheme only relies on linear optics and post-selection algorithm. We discuss the potential of using this amplifier as a building block in extending the distance of quantum communication.

\end{abstract}

\today

\maketitle

\section{Introduction}
\label{sec:Introduction}

The question of quantum noise in linear amplifiers has stirred considerable interest not only because of its technical significance, but also owing to its intimate connection with the most fundamental features of quantum theory. A {\it perfect linear amplifier} (PLA) increases the power of an incoming signal without introducing a degradation to its signal-to-noise ratio (SNR). This is achievable easily for classical signals. However, in the quantum world, a PLA cannot function deterministically. Due to the bosonic nature of photons, an optical amplifier unavoidably introduces noise to any signal it processes. The noise penalty arises from the interaction between the initially independent input mode and the internal modes of an amplifier. This quantum property of amplifiers was theoretically elucidated by Haus and Mullen~\cite{PR128.2407} and was quantitatively expressed as the {\it amplifier uncertainty principle}~\cite{PRD261817}. In particular, for a phase-insensitive amplifier, the minimum amount of additional noise is equivalent to $|G-1|$ units of vacuum noise, where $G$ denotes the power gain for the input signal. This noise penalty prevents the increase of distinguishability of quantum states under amplification. It therefore ensures that by means of the {\it amplify-and-split} approach~\cite{PRA86010305}, two orthogonal quadrature amplitudes of a bosonic mode 
cannot be measured simultaneously with arbitrary precision, in compliance with the Heisenberg uncertainty principle.
 
One way to circumvent the excess noise is to instead apply phase-sensitive amplification. One such example is to squeeze either the input mode, or the internal mode, such that the amplified output has reduced noise in one quadrature at the expense of degrading the conjugate quadrature~\cite{PRA29.1419,PRA35.4443}. Besides, phase-insensitive amplification can also be realised using a series of light emitter detectors in conjunction with high-quantum-efficiency photodetectors~\cite{OL12.922}. This device can achieve, in principle, a signal transfer limited only by the photodetector efficiency ($\mathrm{SNR}_{\mathrm{out}}/\mathrm{SNR}_{\mathrm{in}}{\approx}\eta_{d}$, where $\eta_d$ is the quantum efficiency of the photodetector) for a sufficiently large number of emitters. However, while the intensity of light is amplified, all phase information is destroyed. Another method of low-noise amplification is to use an electro-optic feedforward loop~\cite{PRL79.1471}. The setup avoids the requirement of nonlinear optical process, and due to the fact that not all of the input light is destroyed, some of the phase information can be retained. 

\begin{figure}[!ht]
\includegraphics[width=0.45\textwidth]{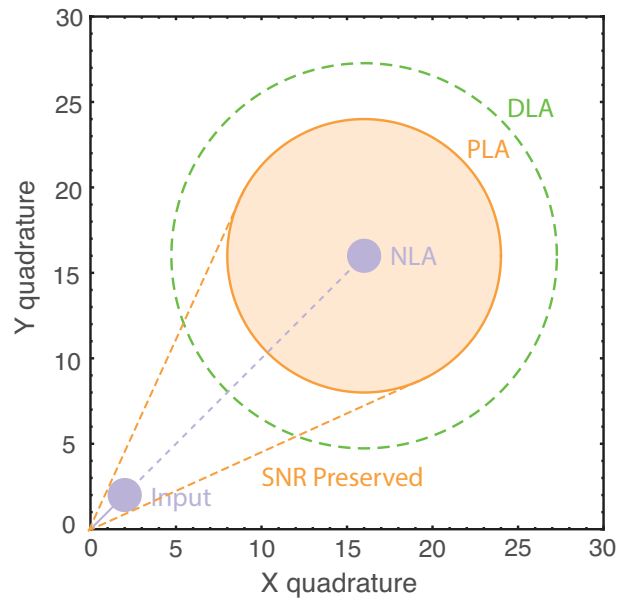}
\caption{\label{fig0} Wigner function contours of input and output coherent states for a continuum of linear amplifiers. The green dashed circle here refers to the best possible deterministic linear amplifier, which adds the minimum amount of noise imposed by quantum mechanics; any amplifier that introduces less noise is necessarily probabilistic. One example of the probabilistic amplifiers is the PLA that preserves the SNR of an incoming signal while amplifying its power (refer to PLA in the graph).  Amplifiers capable of enhancing SNR are called noise-reduced amplifiers (shaded area in orange, including the NLA) and the extreme case of the noise-reduced amplifier is NLA that not only amplifies the amplitude of an input state, but also preserves its noise characteristics. }
\end{figure}

If one demands an amplification of both quadratures equally, an alternative way to evade the noise penalty is to allow a probabilistic operation. Fiur\'{a}\u{s}ek proposed a probabilistic amplification method that could be applied to coherent states of fixed amplitude but unknown phase~\cite{PRA70032308}. Ralph and Lund extended this idea and proposed independently the noiseless linear amplifier (NLA) \cite{Ralph2009} that could in principle be applied to arbitrary ensembles. This amplifier outperforms the perfect linear amplifier by preserving the noise characteristic of the input state and is hence, from a classical point of view, a noise-reduced amplifier, as illustrated in Fig.~\ref{fig0}.  The price to pay is that the process has to be probabilistic and approximate in terms of the output state produced. A better approximation is attainable at the expense of a lower success probability~\cite{NJP15.073014,Ralph2009}. This compromise guarantees that, on average, the Heisenberg's uncertainty relation remains
satisfied. Nevertheless, the successfully amplified quantum states can be
heralded and thus is valuable in extending the range of loss-sensitive
protocols.

Various physical implementations of NLA have been proposed and experimentally demonstrated,  including the quantum scissor setup~\cite{Xiang2010,Ferreyrol2010,Ferreyrol2011,SKocsis2013}, the photon-addition and -subtraction~\cite{Zavatta2010,NatPhoton097642015}, and noise addition~\cite{Usuga2010} schemes. In all these approaches, a large truncation is often imposed on the unbounded amplification operator in the photon-number basis. The high-fidelity operating region of the amplification is consequently restricted to small input amplitude and small gains~\cite{PRA88033852,PRA93052310}. 
The current realisations require non-classical light sources and non-Gaussian operations like photon counting, thereby rendering their application to many systems and protocols very challenging. Intriguingly, as recently proposed~\cite{PRA86060302,PRA87020303} and experimentally demonstrated~\cite{Chrzanowski2014}, the benefits of noise-reduced amplification can be retained via classical post-processing, provided that the NLA  precedes a dual homodyne measurement directly. 
Although the simplicity of this measurement-based noiseless linear amplifier (MB-NLA) is appealing, its post-selective nature confines it to point-to-point applications such as quantum key distribution. To overcome this drawback, the concatenation of an NLA and a deterministic linear amplifier (DLA) that uses MB-NLA and yet outputs a quantum state was proposed recently and studied in the context of quantum cloning \cite{JYHaw2017}. 

In the current paper, we realise a quantum enhancement of signal-to-noise-ratio for arbitrary coherent states using a heralded noise-reduced linear amplifier. This amplifier combines the advantages of a DLA and an MB-NLA. Owing to the fully tunable cutoffs and independent control of the NLA and DLA gains, great versatility in the effective gain and the input amplitude is attained, mitigating therefore the undesirable constraints in previous physical implementations. We show a signal transfer of 110\% from input to output with an amplification gain of 6.18 when Gaussian statistics is maintained. Furthermore, by marginally compromising the Gaussianity of the output state, we demonstrate an SNR enhancement of more than 4dB for a coherent state amplitude of $|\alpha|{=}0.5$ with an amplification gain of 10.54.
Unlike the previous measurement-based NLA scheme~\cite{Chrzanowski2014}, a heralded and free-propagating amplified state is produced with our amplifier. It is worth stressing that the setup uses Gaussian elements and a post-selection algorithm only, and hence has a better compatibility with other continuous-variable protocols. 

\section{Theory}
\label{sec:Setup}
\subsection{Conceptual Scheme}
\begin{figure}[h!]                                                                                                                                                                                                                                                                                                                                                                                                                                                                                                                                                                                                                                                                                                                                                                                                                                                                                                                                                                                                                                                                                                                                                                                                                                                                                                                                                                                                                                                                                                                                                                                                                                                                                                                                                                                                                                                                                                                                                                                                                                                                                                                                                                                                                                                                                                                                              
\centering
\includegraphics[width=0.45\textwidth]{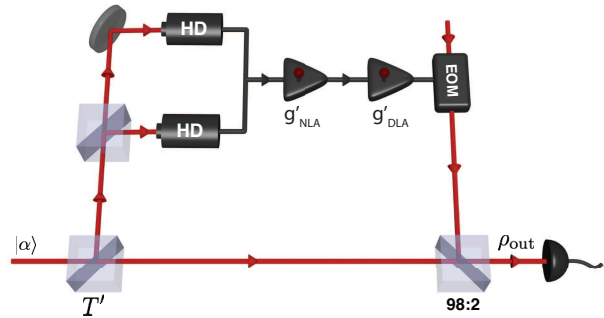}
\caption{\label{fig1} Experimental schematic of our heralded noise-reduced linear amplifier achieved with a feedforward loop. The amplifier has two control parameters, the DLA gain $g_{\mathrm{DLA}}$ and the NLA gain $g_{\mathrm{NLA}}$. HD: homodyne measurement; EOM: electro-optic modulator.
}
\end{figure}

A conceptual layout of our linear amplifier is depicted in Fig.~\ref{fig1}(a), where the NLA and the DLA are parametrised by their respective gains $g_{\mathrm{NLA}}$ and $g_{\mathrm{DLA}}$. The behaviour of our amplifier is dominated by the interfacing between the two intrinsically different amplifiers. A larger DLA gain would contribute to a higher success probability, but also introduce the noise penalty, while a larger NLA gain is the requisite to attain more increase of SNR, however, at the expense of reducing the success probability.

The effect of our linear amplifier on an unknown input state $\hat{\rho}_{\mathrm{in}}$ is to transform the state as follows: 
\begin{equation}
\hat{\rho}_{\mathrm{out}} = \mathcal{N}\mathrm{Tr}_{v}  \{ \hat{U}_{g_{\mathrm{DLA}}} g_{\mathrm{NLA}}^{\hat{n}}  \hat{\rho}_{\mathrm{in}} \otimes  \ket{0}\bra{0}_{v} g_{\mathrm{NLA}}^{\hat{n}} \hat{U}_{g_{\mathrm{DLA}}}^{\dagger} \} \;,
\end{equation}
where the constant \(\mathcal{N}\) is a normalisation factor. The operator \(g_{\mathrm{NLA}}^{\hat{n}} \) here models the action of the NLA on the input density operator, whilst \(\hat{U}_{g_{\mathrm{DLA}}} = e^{-\theta\left(\hat{a}\hat{a}_v-\hat{a}^\dagger \hat{a}_v^\dagger \right) } \) is a unitary transformation acting on the input mode and an ancillary vacuum mode which models the action of the DLA. The parameter \(\theta\) relates to the gain of the DLA via \(g_{\mathrm{DLA}}= \mathrm{cosh}(\theta)\). The ancilla mode is traced out to give the final output. We can characterise the outcome of this interaction by considering the expectation value of an observable \(\hat M(\hat a, \hat a^\dagger)\),
\begin{align}
\langle \hat{M} \rangle &= \mathrm{Tr} \{ \hat{M} \hat{\rho}_{\mathrm{out}} \} \nonumber
\\
&=  \mathrm{Tr} \{ \hat{M} \hat{U}_{g_{\mathrm{DLA}}} g_{\mathrm{NLA}}^{\hat{n}}  \hat{\rho}_{\mathrm{in}} \otimes  \ket{0}\bra{0}_{v} g_{\mathrm{NLA}}^{\hat{n}} \hat{U}_{g_{\mathrm{DLA}}}^{\dagger}\} \nonumber
\\
&=  \mathrm{Tr} \{\hat{M}_{\mathrm{DLA}}   \hat{\rho}_{\mathrm{NLA}}  \}\;,
\end{align}
where we use the cyclic permutation of the trace and \(\hat{M}_{\mathrm{DLA}} =\hat{U}_{g_{\mathrm{DLA}}}^{\dagger} \hat{M} \hat{U}_{g_{\mathrm{DLA}}}\), \(  \hat{\rho}_{\mathrm{NLA}}=  g_{\mathrm{NLA}}^{\hat{n}}  \hat{\rho}_{\mathrm{in}}  g_{\mathrm{NLA}}^{\hat{n}}\).

We first consider the input \(\hat{\rho}_{\mathrm{in}}\) to be an ensemble comprised of a Gaussian distribution of coherent states:
\begin{equation}
 \hat{\rho}_{\mathrm{in}} \left(\lambda\right) = \frac{1}{\pi} \frac{1-\lambda^2}{\lambda^2} \int \mathrm{d}^2 \alpha e^{-\frac{1-\lambda^2}{\lambda^2}|\alpha|^2} \ket{\alpha}\bra{\alpha}
\end{equation}
where \(\lambda\) ($0{\leq}\lambda{<1}$) relates to the variance of the distribution by \(V=\frac{1+\lambda^2}{1-\lambda^2}\). Due to the linearity of the NLA operator, the distribution \(\hat{\rho}_{\mathrm{in}}\left(\lambda \right)\) changes as \(g_{\mathrm{NLA}}^{\hat{n}} \hat{\rho}_{\mathrm{in}}\left(\lambda \right)  g_{\mathrm{NLA}}^{\hat{n}} \propto \hat{\rho}\left(g_{\mathrm{NLA}}\lambda\right)\) under noiseless linear amplification~\cite{PRA86012327}. That is, if Alice sends a distribution of coherent states of width \(\lambda\), the conditional state after the successful operation of NLA is proportional to a distribution of width \(g_{\mathrm{NLA}}\lambda\).  Correspondingly, the variance of the ensemble of coherent states becomes \(V=\frac{1+g_{\mathrm{NLA}}^2\lambda^2}{1-g_{\mathrm{NLA}}^2\lambda^2}\). We note that for the amplified distribution to be physical, \(g_{\mathrm{NLA}}^2\lambda^2\) must be less than one. The state \(\hat{\rho}_{\mathrm{NLA}}\left(g_{\mathrm{NLA}}\lambda\right)\) is then amplified by the DLA to give the final output state. The expectation value of an arbitrary observable \(M(\hat{a}_{\mathrm{out}}, \hat{a}_{\mathrm{out}}^\dagger)\) can then be constructed using~\cite{PRD261817, PRL96163602}
\begin{align}
 \label{eq:dla}
\hat{a}_{\mathrm{out}} &=\hat{U}_{g_{\mathrm{DLA}}}^{\dagger} \hat{a}_{\mathrm{in}} \hat{U}_{g_{\mathrm{DLA}}} \nonumber \\
& = \hat{a}_{\mathrm{in}}\, g_{\mathrm{DLA}} +\hat{a}_v^\dagger  \sqrt{g_{\mathrm{DLA}}^2-1}\;,
\end{align}
So far, we have described how an ensemble of coherent states evolves by our amplifier. We now examine the action of our amplifier on each individual coherent state \(\ket{\alpha}\).

The NLA probabilistically amplifies the complex amplitude of an input coherent state $\ket{\alpha}$ to $\ket{g_{\mathrm{NLA}}\alpha}$ with a gain $g_{\mathrm{NLA}}{>}1$. The DLA then performs the deterministic transformation as shown in Eq.(\ref{eq:dla}). The mean of the amplitude $\hat{X}_{+}=\hat{a}+\hat{a}^{\dagger}$ and phase $\hat{X}_{-}=-i(\hat{a}-\hat{a}^{\dagger})$ quadratures of the electric field is therefore amplified by 
\begin{align}
 \label{eq:theom}
 \langle \hat{X}_{\pm}\rangle_{\mathrm{out}} = g_{\mathrm{NLA}}g_{\mathrm{DLA}} \langle \hat{X}_{\pm}\rangle_{\mathrm{in}}\;.
\end{align}
To quantify the amplification of the signal, we define $g_{\mathrm{eff}}=g_{\mathrm{NLA}}g_{\mathrm{DLA}}$ as the effective gain. Since the NLA incurs no additional noise, the overall output noise is only a function of the DLA gain (where the quantum noise level is 1)
\begin{align}
 \label{eq:theov}
   \langle (\delta \hat{X}_{\pm})^2\rangle_{\mathrm{out}} = 2g_{\mathrm{DLA}}^2-1\;.
\end{align}

\subsection{Equivalent Experimental Scheme}

Figure~\ref{fig1} plots the experimental scheme of the amplifier, where an input coherent state is first fed through a beam splitter with a transmissivity of (see more details in Appendix A)
\begin{equation}
  \label{eq:T'}
  T' = g_{\mathrm{NLA}}^2/g_{\mathrm{DLA}}^2\;.
\end{equation}
The reflected mode is then subjected to a dual-homodyne setup locked to simultaneously measure two conjugate quadratures. An MB-NLA, consisting of a filter function and a rescaling factor, amplifies the mean of the measured statistics by $g'_{\mathrm{NLA}}$ without changing its noise feature. More specifically, a probabilistic Gaussian filter given by
\begin{equation}
  \label{eq:filter}
P(\alpha_m) =
\begin{cases} \exp\left(| \alpha_{\mathrm{m}} |^2{-}| \alpha_{\mathrm{c}} |^2\right) \left(1-\frac{1}{(g'_{\mathrm{NLA}})^{2}} \right)  &\mbox{if } |\alpha_{\mathrm{m}}|{ \leq}
  \alpha_{\mathrm{c}}\\
 1&\mbox{otherwise}
\end{cases}\;
\end{equation} 
is applied to the measurement outcomes $\alpha_{\mathrm{m}}{=}(x_{\mathrm{m}}+ip_{\mathrm{m}})/\sqrt{2}$ of the dual-homodyne station. The cutoff parameter $\alpha_{\mathrm{c}} > 0$ acts as the truncation on the working phase space of the unbounded amplification operator, which therefore determines how closely
the filter approximates an ideal NLA and also the success probability of
the protocol. This filter function heralds the successful amplification and over-amplifies both the mean and the variance of the measured statistics by ${g'_{\mathrm{NLA}}}^{2}$. Thus, to retrieve the target mean and eliminate the additional noise, a rescaling factor of $1/g_{\mathrm{NLA}'}$ is applied to the filtered statistics. The entire functionality of the measurement-based NLA is symbolised as the tunable gain $g'_{\mathrm{NLA}}$ in Fig.~\ref{fig1}(b). The output signal of the measurement-based NLA is further amplified electronically by $g'_{\mathrm{DLA}} = \sqrt{2\left( g_{\mathrm{DLA}}^2-1 \right)}$ and coupled to the transmitted input beam to fulfill the displacement operation (the relationships between $g'_{\mathrm{NLA}}$, $g'_{\mathrm{DLA}}$ and $g_{\mathrm{NLA}}$, $g_{\mathrm{DLA}}$ are addressed in more details in Appendix A). 

The output mean and variance of the quadrature amplitudes can be derived as
\begin{gather}
  \label{eq:outmean}
 \langle \hat{X}_{\pm}\rangle_{\mathrm{out}}  = \left(\sqrt{\frac{1-T'}{2}}\, g'_{\mathrm{DLA}}\, g'_{\mathrm{NLA}}+\sqrt{T'} \right)\,\langle \hat{X}_{\pm}\rangle_{\mathrm{in}}, \\
  \langle (\delta \hat{X}_{\pm})^2\rangle_{\mathrm{out}} = 1+(g'_{\mathrm{DLA}})^{2} \;.
 \end{gather}


We quantify the performance of our amplifier by introducing the signal transfer coefficient,
\begin{equation}
\mathcal{T}_s = \mathrm{SNR}_{\mathrm{out}}/\mathrm{SNR}_{\mathrm{in}}\;,
\end{equation} 
which is equal to
\begin{equation}
  \label{eq:ILA}
g_{\mathrm{eff}}^2/(2g_{\mathrm{DLA}}^2-1)\;
\end{equation}
for a quantum-limited amplification \cite{PRL96163602} and is larger than 1 for a noise-reduced operation. From Eq.~(\ref{eq:outmean}) and (10), we obtain the theoretical $\mathcal{T}_s$ for both quadratures for our setup, 
\begin{align}
 \label{eq:Tsid}
\mathcal{T}_s  = \frac{(\sqrt{\frac{1-T'}{2}}\, g'_{\mathrm{DLA}}\, g'_{\mathrm{NLA}}+\sqrt{T'})^2}{1+(g'_{\mathrm{DLA}})^{2}}\;.
\end{align}
Another performance metric--the success probability--is also calculated so as to define the operating region where the amplification is experimentally feasible (see Appendix B).

The key features of our hybrid linear amplifier are threefold: first, the output is a free-propagating amplified physical state; second, the setup only depends on linear optics; third, the two cascading gains can be tuned independently and our cutoff is fully adjustable. This introduces more flexibility in optimising the success rate while preserving high fidelity with an ideal implementation of NLA. It also largely extends the operating region of the amplifier by alleviating the constraints of previous physical implementations where amplification is confined to small input amplitudes and low amplification gains. 

\begin{figure}[!ht]
\centering
\includegraphics[width=0.45\textwidth]{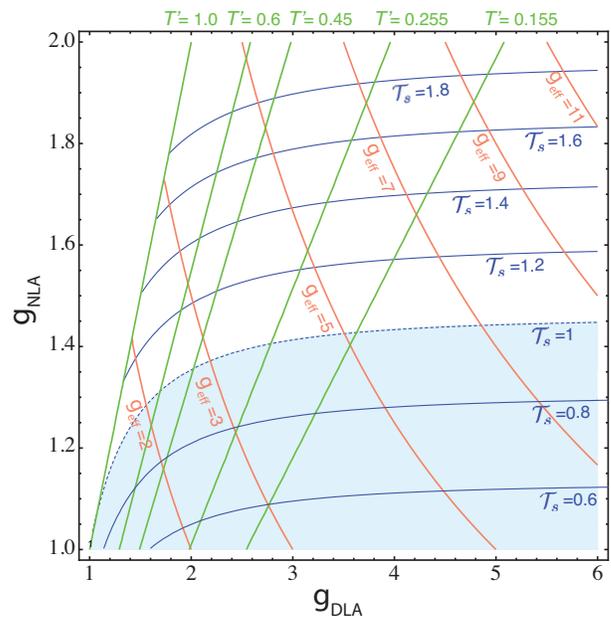}
\caption{Tunability of the amplifier. 
Signal transfer coefficient (blue contours), various effective gains (red contours), and $T'$ (green lines) as the function of $g_{\mathrm{NLA}}$ and $g_{\mathrm{DLA}}$. The blue-dotted line denotes the amplification process where the input SNR is preserved, while the enclosed shaded area refers to the region where additional noise is introduced. We note that, without a sufficiently high NLA gain, increasing $g_{\mathrm{DLA}}$ alone would not suffice to approach the noise-reduced amplification.
}
  \label{fig:flex}
\end{figure}

Figure~\ref{fig:flex} illustrates the operational degrees of freedom of our noise-reduced linear amplifier. The amount of noise reduction depends on both the product and the ratio of $g_{\mathrm{NLA}}$ and $g_{\mathrm{DLA}}$, which correspond to, respectively,  the values of the effective gain $g_{\mathrm{eff}}$ and the transmittivity $T'$ in Fig.~(\ref{fig1}). Intuitively, for a fixed effective gain $g_{\mathrm{eff}}$, a higher signal transfer coefficient $\mathcal{T}_s$ becomes more pronounced with a larger $g_{\mathrm{NLA}}$, since the associated noise determined by $g_{\mathrm{DLA}}$ decreases while the input amplitude undergoes the same amount of amplification. Hence, under the same effective gain, a higher $T'$ would always lead to a larger signal transfer coefficient. 

We note that there is an ultimate limit of our current setup embodied in Eq.~(\ref{eq:T'}). Because $T'<1$, $g_{\mathrm{NLA}}$ must be smaller than $g_{\mathrm{DLA}}$. This, in terms of the effective gain, poses a limit on the signal transfer as
\begin{equation}
  \label{eq:5}
\mathcal{T}_s< \frac{g_{\mathrm{eff}}^2}{2 g_{\mathrm{eff}}-1}\;
\end{equation}
(see Appendix A).  Nonetheless, as shown in Fig.~\ref{fig:flex}, an arbitrarily high $\mathcal{T}_s>1$ is attainable using the current setup by applying well-tailored $T'$ and $g_{\mathrm{eff}}$.

\section{Experimental setup}
\label{sec:Experiment}
The light source for this experiment is an Nd:YAG laser producing continuous wave singe mode light at 1064 nm. The coherent state at a sideband frequency is generated by sending modulation signals at \SI{4}{MHz} to a pair of electro-optical modulators (EOMs) on the signal beam. The laser was found to be shot-noise-limited at this frequency and the amplitudes of the modulation signals determine the complex amplitude of the coherent state. To amplify the coherent state, we first inject the input state into a beam splitter with transmittivity of $T'$ where it is split to the transmitted and reflected modes. A dual homodyne measurement is then performed on the reflected mode and the measurement outcomes serve two purposes. First, they are used to extract the 4-MHz modulation and to reveal the term $|\alpha_m|$ in Eq.~(\ref{eq:filter}) which is used to provide the heralding signal. To this end, the outcome is demodulated by mixing it with an electronic local oscillator, before being low pass filtered at \SI{100}{kHz} and oversampled on a 12-bit analog-to-digital converter at \SI{625}{kSa} per second.
Second, the outcomes of the dual homodyne measurement are also employed to accomplish the feedforwarding. They are amplified
electronically with a gain $g_{\mathrm{ele}}{=}g'_{\mathrm{DLA}}/g'_{\mathrm{NLA}}$ and fed into
a pair of EOMs modulating a bright
auxiliary beam. This intense beam is then coupled in phase with the transmitted signal beam by an asymmetric beam splitter of transmissivity 98\% to realise the displacement operation. 

The combined beam is then characterised by a homodyne
measurement, locked alternatively to amplitude and phase quadratures. The homodyne measurement goes through the same signal processing and at least $5\times10^7$ data points are acquired.

\section{Results}
\label{sec:Results}

\subsection{LINEARITY OF THE AMPLIFIER}
\label{sec:linear}
\begin{figure*}[h!]
 \begin{center}
\includegraphics[width=1.0\textwidth]{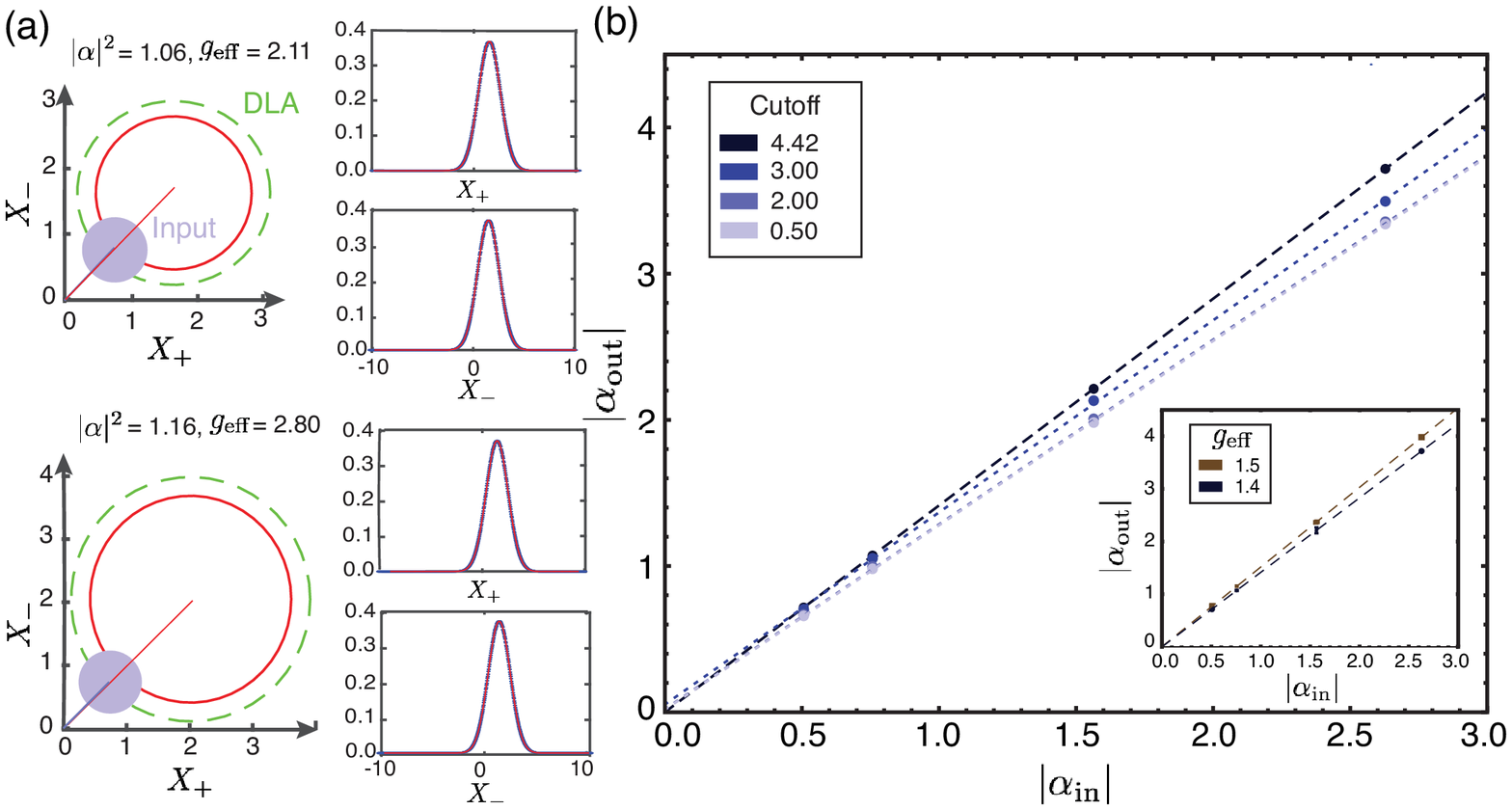}
\caption{Linearity of the amplifier. (a) Amplification for coherent states with different amplitudes. Left panels: noise contours (one standard deviation width) of the amplified states. Right panels: normalised probability distribution for amplitude and phase quadratures of the output states. (b) Output magnitudes vs input magnitudes as we reduce the cutoff whilst maintaining the values of $g_{\mathrm{NLA}'}$ and $g_{\mathrm{DLA}'}$. Inset: Output magnitudes vs input magnitudes with cutoff being $\alpha_c=4.42$ at different effective gains.
}
\label{fig:linear}
\end{center}
\end{figure*}

Figure~\ref{fig:linear} shows the performance of our linear amplifier for input coherent states with different complex amplitudes $|(x+ip)/2\rangle$. As illustrated in Fig.~\ref{fig:linear}(a), we demonsrate the phase-preserving property of the amplifier, and observe symmetric noise spectrum of amplitude and phase quadratures.
These results emerge from the linearity and phase invariance of the present setup, as is also clearly demonstrated in Fig.~\ref{fig:linear}(b). In particular, under the same $T'$ (0.6), by selecting input states with different complex amplitudes $(x,p)=(-0.71,0.72),(-0.01,-1.51),(2.23,2.19),(5.26,-0.02)$, we plot the output magnitudes against the input magnitudes as we vary the cutoff values, or alternatively as we vary the effective gains. The amplifier behaves linearly in either circumstance, thus verifying the independence of the amplification on the input states. 

Apart from the relationship between $|\alpha_{\mathrm{out}}|$ and $|\alpha_{\mathrm{in}}|$ shown in Fig.~\ref{fig:linear}, we also notice that as we reduce the cutoff, the output states start to exhibit non-uniform noise between the in-phase and out-of-phase fluctuations. More specifically, as the cutoff is decreased from 4.42 to 0.50, we observe the output noise $[ \langle (\delta \hat{X}_{+})^2\rangle_{\mathrm{out}},\langle(\delta \hat{X}_{-})^2\rangle_{\mathrm{out}}]$ reduces, respectively, from $[1.83,1.83]$ to $[1.59,1.70]$ for input $(x,p)=(-0.01,-1.51)$, and from $[1.87,1.86]$ to $[1.70,1.58]$ for input $(x,p)=(5.26,-0.02)$. In these cases, the cutoff with respect to the effective gain no longer suffices to preserve the Gaussianity of the output state and the amplified states start to squash along the radial direction \cite{PRA88033852}. 
Nevertheless, the amplification remains phase insensitive due to the fact that it is always the variance of the quadrature along the radial direction that becomes classically "squeezed", whilst that of the orthogonal quadrature inclines to be anti-squeezed. Interestingly, it is worth emphasising that, even in this operating region, the amplifier still works equivalently for all input amplitudes regardless of the insufficient cutoff (refer to the light blue line in Fig. \ref{fig:linear}(b)). This special property would be of great benefit for coherent states discrimination. For example, we consider multiple weak coherent states, in a quadrature phase-shift-keyed format \cite{RMP84621}, as inputs of our linear amplifier. Regardless of the phases of the input states, the amplifier increases their complex amplitudes consistently and, meanwhile, suppresses the added noise along the radial direction. The amplification works conditionally, whereas as long as a heralding signal reveals that the amplifications succeed for all  input states, the distinguishability of these states would be enhanced. 

\subsection{QUADRATURE INDEPENDENCE AND THE SUCCESS PROBABILITY}
\label{sec:versatility}
\begin{figure*}[!ht]
\centering
\includegraphics[width=0.8\textwidth]{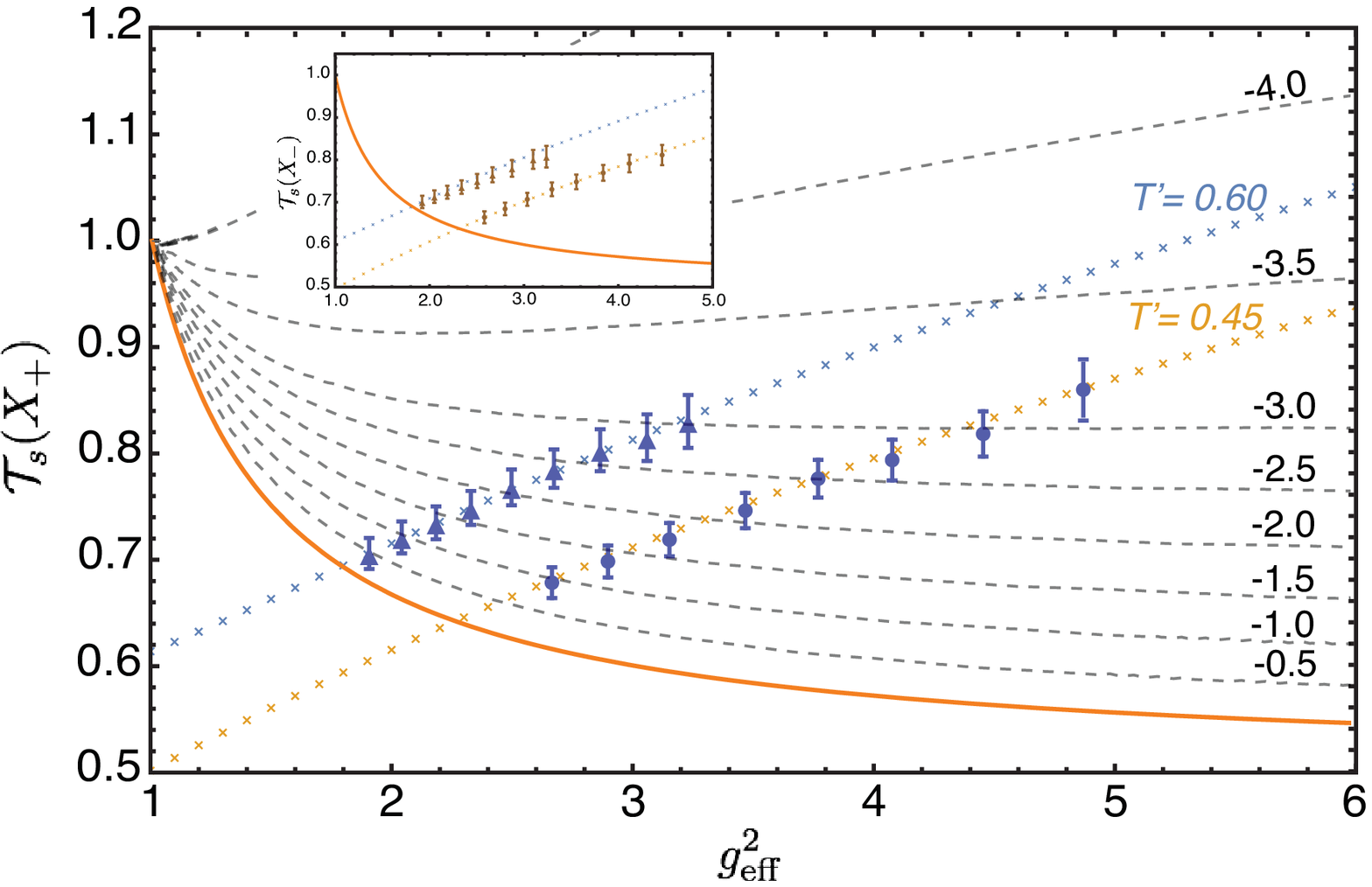}
\caption{Amplifier performance: noise properties and probability. Signal transfer coefficient for amplitude quadrature (blue symbols) as a function of $g_{\mathrm{eff}}^2$ for varying $T'$: 0.6, 0.45. The theoretical prediction, assuming infinite cutoff, is depicted in crosses. It is clearly shown that the experimental $\mathcal{T}_s$ increases in compliance with the prediction, demonstrating that the cutoff ($\alpha_c=4.3$) selected is sufficient and no over- or under-estimation of $\mathcal{T}_s$ appears. The success probability in logarithmic scale is superimposed which decreases as we increase $g_{\mathrm{eff}}$. For the sake of comparison, the best achievable $\mathcal{T}_s$ of an optimal deterministic linear amplifier (also termed as the quantum noise limit) is shown in orange solid line. Inset: the experimental data superimposed with its theoretical prediction for phase quadrature.
}
\label{fig:Probability}
\end{figure*}

Figure~\ref{fig:Probability} demonstrates the tunability and versatility of our hybrid linear amplifier. The signal transfer coefficients of the amplitude and phase quadratures, superimposed by the success probability, are plotted as a function of increasing effective gains. We examine an input coherent state with complex amplitude of $(x,p)=(1.51,1.54)$ for all plots. Two different transmissions, $T'=0.60$, and 0.45, are picked to test the amplifier in different settings. 

In accordance with Fig.~\ref{fig:flex}, data points with the same $T'$ illustrate evidently the improving of $\mathcal{T}_s$ as $g_{\mathrm{eff}}$ increases, corresponding to moving along the green lines in Fig.~\ref{fig:flex}. Alternatively, when keeping $\mathcal{T}_s$ constant, lowering down $T'$ results in a smaller success probability, which also coincides with Fig.~\ref{fig:flex}, because this decrease of the success probability results from the increase of $g_{\mathrm{NLA}}$.

We note that all $\mathcal{T}_s$, for both amplitude and phase quadratures, exceed the quantum limit regardless of the values of $T'$, among which the maximum achieved $\mathcal{T}_s$ are $0.830\pm0.025$ and $0.860\pm0.024$ for $T'=0.6$, and $T'=0.45$, respectively. These results significantly surpass the maximum allowable signal transfer in the deterministic regime (c.f.\ Eq.~(\ref{eq:ILA})) by around 10 and 12 standard deviations, respectively. All the observed values of $\mathcal{T}_s$ show good agreement with the theoretical model assuming infinite cutoff and taking into account the experimental imperfections (see Appendix C). The corresponding success probability ranges between \num{e-3} and 0.3, rendering the amplifier still relatively practicable. Slight discrepancies are observed between $X_{+}$ and $X_{-}$ owing to the different losses experienced by the two quadratures during feedforwarding (see  Appendix D). Small deviations of the experimental data from the prediction are attributed to other in-line electronic noise.

\subsection{HIGH SIGNAL TRANSFER COEFFICIENTS}
\label{sec:highT}

\begin{figure*}[!ht]
\centering
\includegraphics[width=0.6\textwidth]{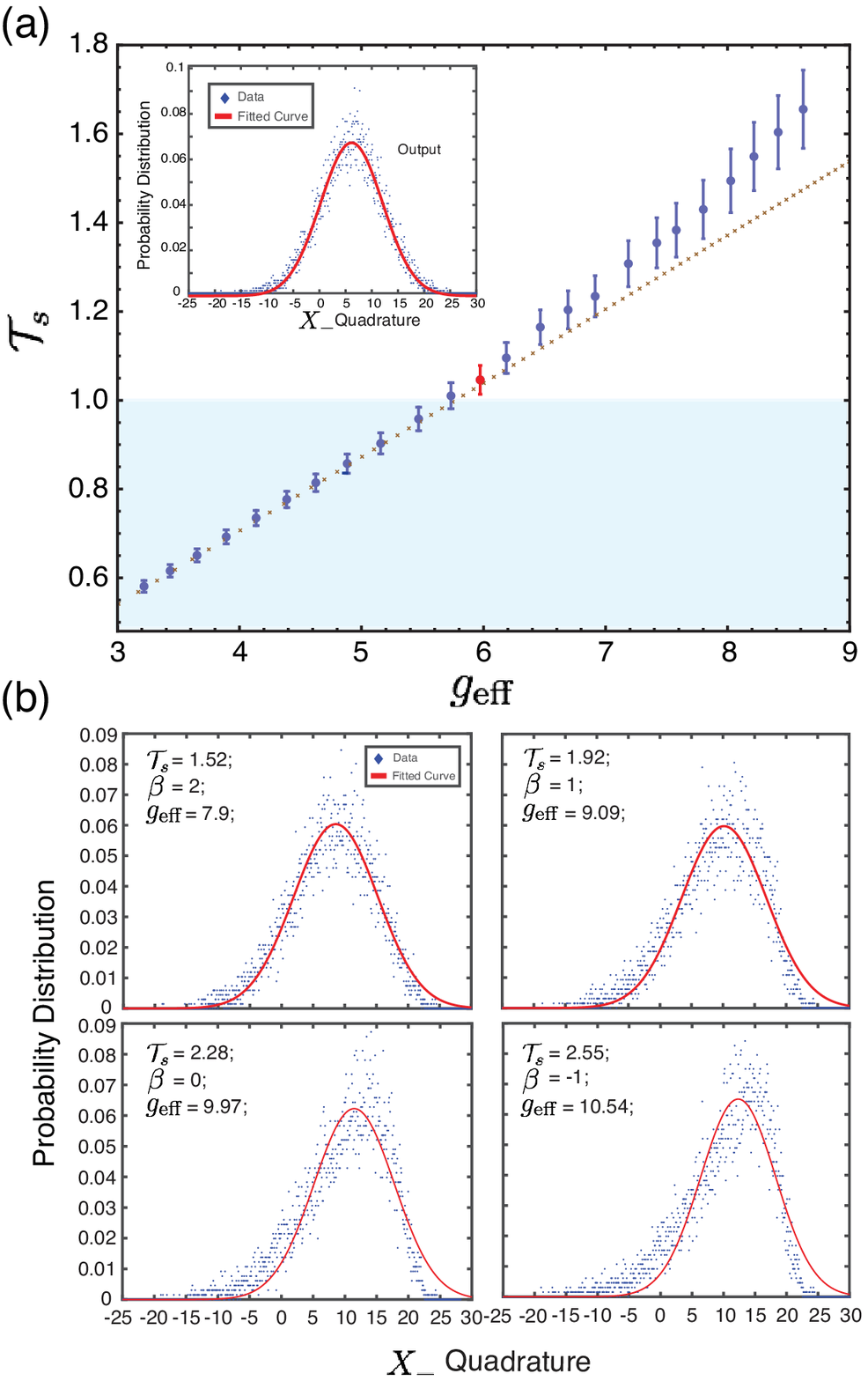}
\caption{Signal transfer coefficients in the large gain domain. (a) $\mathcal{T}_s$ exceeding 1 with increasing $g_{\mathrm{eff}}$ for $\alpha_c=4.5$ and a coherent state amplitude of $|\alpha|=0.5$. The experimental $\mathcal{T}_s$ shows good agreement with the theory plot (in crosses) until around $g_{\mathrm{eff}}=6.5$ where the data points start to depart, thereby indicating that the cutoff no long suffices to maintain the output Gaussianity. Inset: probability distribution of the amplified state labeled in red.  (b) Probability distributions of the phase quadrature of the amplified state with cutoffs given by $\alpha_c=g_{\mathrm{NLA}'}^2 |\alpha_{\mathrm{in}}|+\beta \sqrt{0.5} g_{\mathrm{NLA}'}$. The parameter $\beta$ quantifies how 
well is the cutoff circle able to embrace the distribution of the amplified state. Data points are the post-selected ensemble out from $2.7\times10^9$ homodyne measurements while the red curves indicate the corresponding best-fitted Gaussian distributions.  }
\label{fig:LargeNF}
\end{figure*}

In Fig.~\ref{fig:LargeNF}, we summarise our experimental results when our amplifier is operating in the large gain domain for an input state $(x,p){=}(0,1.01)$($|\alpha|{=}0.5$) and $T'=0.155$. As is shown in Fig.~\ref{fig:LargeNF}(a), a higher $\mathcal{T}_s$ is obtained at the expense of a lower success probability. 
We see that the increasing of $\mathcal{T}_s$ as a function of $g_{\mathrm{eff}}$ coincides with the theoretical model based on an infinite cutoff (see Appendix C), indicating that the cutoff employed is sufficient to encompass the amplified distribution and thus exclude any distortion of the output. In this high-fidelity operating region, a $\mathcal{T}_s$ larger than 1 (specifically, $1.10\pm0.04$) is observed, thus verifying a clear fulfillment of the noise-reduced amplification. As $g_{\mathrm{eff}}$ keeps increasing, a wider discrepancy appears between the experimental value of $\mathcal{T}_s$ and its theoretical prediction, as the result of an insufficient cutoff. In this situation, to maintain a Gaussian statistics output, more data points are required.

To complete the investigation of our setup, we also explore the relationship between $\mathcal{T}_s$ and the output Gaussianity while keeping the success probability unchanged (around $10^{-6}$), as shown in Fig.~\ref{fig:LargeNF}(b). In this case, as we relax the requirement for the output Gaussianity, it is possible to enjoy a higher effective gain and therefore achieve a considerably larger $\mathcal{T}_s$ without decreasing the success probability. We experimentally obtained an signal transfer of $\mathcal{T}_s{=}2.55\pm0.076$ from input to output with an amplification gain of 10.54. 

\section{Discussions and conclusions}
\label{sec:Discussions}

In this paper, we demonstrate an enhancement of signal-to-noise ratio for arbitrary coherent states with a noise-reduced linear amplifier that profitably combines a measurement-based noiseless linear amplifier and a determnistic linear amplifier. We also investigate the possibility of applying our amplifier to an ensemble of coherent states. The hybrid nature of the amplifier retains the flexible and operational characteristics of the measurement-based NLA, which, as opposed to the physical implementations, evades the demand of nonclassical light sources, and the restriction to small input states and low amplification gains. 
It also preserves the free-propagating amplified states and thus circumvents the drawback of a pure measurement-based setup whose output can only be classical statistics. Even though the amplifier works conditionally, a heralding signal is generated for successful events. We observe a signal transfer coefficient $\mathcal{T}_s$ larger than 1, clearly showing that the amplification is noise-reduced. We also demonstrate that higher $\mathcal{T}_s$ -- more specifically, $\mathcal{T}_s=2.56$ with $g_{\mathrm{eff}}=10.54$ -- is attainable if one is willing to accept a lower success probability or instead to compromise slightly the output Gaussianity. Interestingly, we also notice that there exists an operating region where the amplifier works linearly, regardless of the relatively small distortion of the output. This would provide a useful coherent state discrimination machine. Owing to the composability, tunability, and ease of implementation of our amplifier, we expect it to have applications in a broad range of quantum information protocols, including entanglement distillation \cite{Xiang2010, NatPhoton097642015, Chrzanowski2014}, quantum cryptography \cite{PRL105070501, PRA86012327}, error correction \cite{ TCRalph2011}, quantum teleportation \cite{PRA93012326, PRL115180502} and quantum repeater \cite{PRA95022312}.

\section{Acknowledgement}
We would like to thank G. Guccione and Nelly Huei Ying Ng for their helpful discussions. 
This research is supported by the Australian Research Council (ARC)
under the Centre of Excellence for Quantum Computation and
Communication Technology (project number CE110001027).

\bibliographystyle{unsrt}
 \bibliography{NLA}

\end{document}